\documentclass{article}

\usepackage{PRIMEarxiv}

\usepackage[utf8]{inputenc} 
\usepackage[T1]{fontenc}    
\usepackage{hyperref}       
\usepackage{url}            
\usepackage{booktabs}       
\usepackage{amsfonts}       
\usepackage{amsmath}
\usepackage{nicefrac}       
\usepackage{microtype}      
\usepackage{lipsum}
\usepackage{fancyhdr}       
\usepackage{graphicx}       
\usepackage[authoryear]{natbib}

\title{Relative plausibility versus probabilism: A level-of-analysis error in juridical proof
}

\author{
  Stanley E. Lazic \\
  Prioris.ai Inc. \\
  Ottawa, Canada \\
  \texttt{stan.lazic@cantab.net}
}

\begin{document}
\maketitle

\begin{abstract}
Debates about juridical proof are often framed as a conflict between probabilistic approaches and relative plausibility theory (RPT). This paper argues that this opposition rests on a level-of-analysis error. Drawing on Marr's distinction between levels of analysis, we show that RPT and probabilistic approaches operate at different conceptual levels and are therefore compatible rather than competing theories. RPT provides a computational-level description of juridical proof, characterizing the task of comparing explanations in light of the evidence and assessing whether a standard of proof has been met. Probabilistic approaches supply algorithmic-level accounts that specify how such comparative assessments can be represented and computed. When plausibility judgments satisfy minimal coherence conditions, relative plausibility corresponds to posterior odds. Recognizing this distinction clarifies longstanding disputes and highlights the complementary roles of explanation and probability in legal reasoning.
\end{abstract}


\section{Introduction}

Probabilistic Bayesian inference is often taken as the paradigmatic approach to evidence evaluation in both legal and scientific contexts \citep{Finkelstein1970,Kaplan1968,Good1950,Bovens2003,Bradley2015,Horwich2016}. Nevertheless, dissatisfaction with probabilistic approaches has led to the development of alternatives such as inductive probability \citep{Cohen1977}, belief functions \citep{Clermont2017}, likelihood-based accounts \citep{Sullivan2019}, and various forms of plausible or explanatory reasoning \citep{Rescher1976,McManaman2019}. A prominent recent contribution to this landscape is \emph{relative plausibility theory} (RPT), which has been advanced as a descriptive account of how fact-finders assess evidence and reach verdicts \citep{Pardo2008,Allen2019,Allen2019a,Allen2023}\footnote{\citet{Allen2019a} state that RPT has a wider scope: ``\dots relative plausibility is not only about jury decision-making and the decision rules at trial. Its scope is much broader. It is about the entire process of proof, including (1) the form, securing, and presentation of evidence, (2) the forms of argumentation employed at trial, (3) the manner in which humans process and deliberate on evidence, (4) the trial structure created by the rules of evidence and procedure, (5) the structure of litigation before and after trial, (6) the manner in which judges and juries, on the one hand, and trial and appellate judges, on the other hand, interact, and (7) to some extent, the meaning and nature of rationality. In other words, all of these features comprise the sprawling entity that we refer to as ‘juridical proof’, and relative plausibility is an attempt to explain that entity in all of its aspects, from beginning to end.'' This paper focuses mostly on decision-making and the decision rules}.

At its core, RPT characterizes juridical proof as a comparative process. Competing explanations of the evidence -- typically those advanced by the prosecution or plaintiff and the defence -- are evaluated relative to one another, and a verdict is reached when one explanation is judged sufficiently more plausible than its rival to satisfy the applicable standard of proof, whether beyond a reasonable doubt, the preponderance of evidence, or some other threshold. Importantly, RPT emphasizes that such judgments are not based solely on likelihoods, but also on broader explanatory considerations such as coherence, consistency, simplicity, evidential coverage, consilience, and fit with background knowledge \citep{Allen2019}. The merits and limitations of RPT, and its relationship to probabilistic approaches, have been the subject of extensive debate, including a special issue of this journal (Volume 23 Issue 1--2, April 2019) and subsequent exchanges \citep{Aitken2022,Allen2023}.

This paper argues that much of the debate rests on a \emph{level-of-analysis error}. A level-of-analysis error occurs when descriptions or explanations at different levels of a system or phenomenon are mistakenly treated as competing explanations. Consequently, shortcomings identified at one level are taken as support for an explanation at another level. In the present context, objections to probabilistic methods -- such as their reliance on numerical representations, explicit calculations, or allegedly unrealistic assumptions about human reasoning -- are often interpreted as indirect support for an alternative theory. As a result, theories that address fundamentally different explanatory questions are incorrectly portrayed as rivals. The claim of a level-of-analysis error is not that proponents of RPT misunderstand legal proof, but rather that criticisms of algorithmic representations are frequently treated as evidence favoring computational characterizations of the inferential task.

To make this diagnosis precise, we draw on David Marr's influential distinction between levels of analysis \citep{Marr2010}. Any system that performs an information-processing task can be described at three levels. At the \emph{computational} level, a theory specifies what problem the system solves, why it solves it, and how inputs are related to outputs. At the \emph{algorithmic} level, a theory specifies how the task is carried out: what representations are used and what operations manipulate those representations. A third level concerns physical implementation, which we do not consider here\footnote{The implementation level describes how the system is physically realised; for example, as transistors in a circuit board or as neurons in a brain.}.

We argue that relative plausibility theory operates at the computational level. It characterizes the task of juridical proof as one of comparing competing explanations of the evidence and determining whether one explanation is sufficiently more plausible than its alternatives to meet a legally defined standard of proof. Probabilistic approaches, by contrast, operate at the algorithmic level. They specify how comparative assessments of evidential support can be represented, combined, and updated in a coherent manner. Once these distinct explanatory roles are recognized, RPT and probabilistic approaches are no longer rivals but complementary descriptions of the same inferential enterprise.

The importance of this distinction can be illustrated by a simple analogy. Suppose a system receives the set of numbers $\{5,6,1,8,3\}$ and returns $\{1,3,5,6,8\}$. At the computational level, we may describe the system as \emph{sorting} numbers from low to high. At the algorithmic level, however, we must specify how this sorting is achieved; for example, by a bubble-sort procedure \citep{Cormen2009}. Criticizing a particular sorting algorithm as inefficient or implausible does not provide support for the claim that the system sorts numbers. Likewise, showing that jurors do not explicitly perform probabilistic calculations does not provide support for the computational characterization of juridical proof as comparative assessment of evidential support.

Seen in this light, many familiar criticisms of probabilistic approaches lose their force. Claims that fact-finders reason holistically rather than item-by-item, comparatively rather than absolutely, or explanatorily rather than numerically do not refute probabilistic approaches as such. They merely reject particular algorithmic implementations. Rejecting an implementation does not invalidate the computational description of the task itself. \emph{The central claim of this paper is that disputes between relative plausibility theory and probabilistic approaches arise from a level-of-analysis error: they address different levels of explanation and therefore cannot be competing theories of juridical proof.}

The remainder of the paper develops this argument in detail. We first clarify how degrees of plausibility can be represented in a way that respects basic coherence requirements, drawing on foundational results due to Cox \citep{Cox1946,Cox1961,Jaynes2003}. We then show that, under these minimal constraints, relative plausibility corresponds to posterior odds, thereby providing an algorithmic realization of the computational task described by RPT. Finally, we address common claims that probabilistic approaches cannot accommodate explanatory considerations such as fit with background knowledge, simplicity, coherence, or evidential completeness, and explain how these factors can be incorporated at the algorithmic level without departing from the core insights of RPT.

\section{Representing plausibility and coherence requirements}
\label{sec:coherence}

Relative plausibility theory characterizes juridical proof as a task of comparing competing explanations of the evidence. However, if such comparisons are to guide reasoning, verdicts, or standards of proof, degrees of plausibility must be representable in some systematic way. This raises a foundational question: how can subjective assessments of plausibility be represented so that they support coherent reasoning and comparison?

At a minimum, degrees of plausibility must allow for ordering. If one explanation is judged more plausible than another given the same body of evidence, this ordering should be preserved across equivalent states of knowledge. It is therefore natural to represent plausibilities by real numbers, with larger values corresponding to greater plausibility. This choice does not yet commit us to probability theory; it merely provides a convenient and expressive representational framework.

Let $[H \mid E]$ denote the plausibility of a hypothesis $H$ given evidence $E$. Relative plausibility theory is centrally concerned with comparisons of the form
\begin{equation}
   \mathrm{Relative\ plausibility} = \dfrac{[H_P \mid E]}{[H_D \mid E]},
  \label{eq:RP}
\end{equation}

\noindent where $H_P$ and $H_D$ are the explanations advanced by the prosecution and defence, respectively\footnote{$H_D$ is the actual single explanation proposed by the defence, not simply the negation of $H_P$, nor the space of all possible explanations.}. The question is not whether such quantities can be written down, but what constraints they must satisfy if they are to support rational inference.

Following Cox, we may impose a small set of minimal coherence requirements on plausibility assignments \citep{Cox1946,Cox1961,Jaynes2003}. These requirements are not specific to legal reasoning and do not presuppose probabilistic updating; rather, they express general constraints on any system of reasoning that aims to be internally consistent.

First, plausibility assignments should respect the basic rules of propositional logic. If a proposition is true, its negation cannot also be true. Second, plausibility judgments should exhibit qualitative correspondence with common sense: evidence that supports one hypothesis more than another should increase its relative plausibility, irrelevant evidence should have no effect, and the same evidence should not be counted multiple times. Third, reasoning should be path independent: if a conclusion can be reached by different routes; for example, by considering motive before opportunity or vice versa -- those routes should lead to the same result. Fourth, all relevant evidence should be taken into account, subject to admissibility constraints. Finally, equivalent states of knowledge should be represented by equivalent plausibility assignments.

These requirements do not determine a unique numerical scale for plausibility, nor do they specify how evidence must be evaluated in practice. They merely rule out representations that lead to contradictions, order-dependence, or incoherent updates. Importantly, they apply equally to any attempt to formalize the inferential task described by relative plausibility theory. Remaining at the computational level avoids confronting these constraints explicitly, but any algorithmic realization of RPT must ultimately satisfy them. Once plausibility is represented in a way that satisfies these minimal coherence requirements, the question becomes whether such representations admit a systematic and non-arbitrary form. Cox's result shows that they do. Nothing in this result implies that relative plausibility theory must be abandoned or replaced; it shows only that any algorithmic realization of comparative plausibility that satisfies minimal coherence constraints will exhibit probabilistic structure. In other words, anything that satisfies certain reasonable consistency axioms becomes probability theory.

\section{Relative plausibility and posterior odds}
\label{sec:posteriorodds}

Given the coherence requirements outlined above, a natural question arises: can degrees of plausibility be represented in a way that satisfies these constraints while remaining faithful to the comparative task described by relative plausibility theory? Cox's foundational result shows that the answer is affirmative.

Cox demonstrated that any system of plausibility assignments that satisfies basic coherence requirements can be mapped onto the real interval $[0,1]$ in such a way that plausibilities obey the rules of probability theory \citep{Cox1946,Cox1961}. Jaynes later clarified and extended this result, emphasizing that probability theory is not an optional add-on to rational inference, but the unique extension of logic to situations involving uncertainty \citep{Jaynes2003}.

Under such a representation, plausibilities become probabilities, and relative plausibility takes a particularly simple and familiar form. Equation~\eqref{eq:RP} can be rewritten as
\begin{equation}
  \mathrm{Relative\ plausibility} = \dfrac{P(H_P \mid E)}{P(H_D \mid E)},
\end{equation}

\noindent which is simply the posterior odds of the two competing explanations given the evidence \citep{Aitken2022,Aitken2024}. Posterior odds can in turn be expressed as the product of a likelihood ratio (or Bayes factor) and prior odds:

\begin{equation}
  \dfrac{P(H_P \mid E)}{P(H_D \mid E)} =
  \dfrac{P(E \mid H_P)}{P(E \mid H_D)} \times
  \dfrac{P(H_P)}{P(H_D)}.
  \label{eq:BF}
\end{equation}

This result directly addresses a common criticism of relative plausibility theory, namely that it is ambiguous or underspecified with respect to the relationship between plausibility and probability \citep{Gelbach2019,Schwartz2019,Nance2019}. The underspecification is real, but it is not a defect. RPT deliberately operates at the computational level, characterizing the task of evidential comparison without committing to a particular algorithmic implementation. Once the task is instantiated in a way that respects minimal coherence requirements, however, probabilistic structure emerges necessarily.

From this perspective, probabilities are neither competitors to nor parasites upon relative plausibility. They provide one algorithmic realization of the comparative task that RPT describes. If one remains entirely at the computational level, probabilities need not be mentioned at all. But once questions of representation, combination of evidence, and consistency across updates are addressed, probability theory supplies a uniquely coherent framework.

Importantly, this conclusion does not depend on controversial assumptions about human psychology or on the claim that fact-finders explicitly compute probabilities. It follows from the much weaker requirement that plausibility judgments be representable in a way that avoids contradiction, order dependence, and incoherence. Relative plausibility theory and probabilistic reasoning therefore address different explanatory questions: the former specifies what juridical proof aims to achieve, while the latter specifies how that aim can be realized under minimal rational constraints.

\section{Relative probability as a normative algorithmic account}
\label{sec:normative}

The probabilistic representation developed in the previous section is not merely a formal convenience; it has direct normative implications for how comparative evidential reasoning ought to proceed. Once relative plausibility is given an algorithmic realization that satisfies minimal coherence requirements, it naturally takes the form of a comparative probabilistic assessment. This observation helps to resolve a further source of confusion in the literature. Allen and Pardo often characterize probabilistic approaches as non-comparative, contrasting them with the explicitly comparative structure of relative plausibility theory \citep{Allen2019}. However, as shown in Eq.~\eqref{eq:BF}, probabilistic reasoning need not, and typically does not, proceed by assessing hypotheses in isolation. The posterior odds directly compare how well competing explanations account for the evidence. In form and function, they mirror the relative plausibility ratio defined in Eq.~\eqref{eq:RP}. The account offered here is normative and formal rather than psychological; it does not purport to describe how fact-finders actually reason in practice.

The comparative evaluation of hypotheses using odds ratios has a long pedigree in both philosophy and statistics \citep{Peirce2014,Good1950,Good1979,Jaynes2003,Fairfield2022}. The logarithm of this quantity is often referred to as the \emph{weight of evidence}, although that term has acquired multiple and sometimes conflicting meanings \citep{Weed2005}. To avoid this ambiguity and to emphasize continuity with relative plausibility theory, we will refer to posterior odds simply as \emph{relative probabilities}. The term relative probability is used here solely as a descriptive gloss on posterior odds, to emphasize their comparative structure; it does not denote a distinct theoretical framework.

Understood in this way, probabilistic reasoning provides a normative algorithmic account of the inferential task described by RPT. It specifies how a trier of fact \emph{ought} to compare explanations if their judgments are to remain coherent, transparent, and responsive to the totality of the evidence. One influential articulation of this approach is given by \citet{Fairfield2022}. Abstracting from legal detail, the core structure can be summarized as follows.

\begin{enumerate}
\item Begin with a well-defined question, such as ``Who murdered the Colonel?''

\item Specify two or more competing explanations or hypotheses that address the question. In legal contexts, these will typically correspond to the explanations advanced by the prosecution or plaintiff and by the defence. The hypotheses should be mutually exclusive, though they need not be exhaustive. Importantly, the defence must articulate a substantive alternative explanation rather than merely asserting ``not guilty''; for example, that the accused was present but did not commit the act\footnote{See \citet{Fairfield2022} on why this is important.}.

\item Assemble the body of relevant and admissible evidence.

\item Evaluate how well the evidence supports each explanation by assessing the likelihood ratio
\[
\dfrac{P(E \mid H_P)}{P(E \mid H_D)}.
\]
This quantity represents how much more likely the evidence would be if the prosecution's explanation were true rather than the defence's. Evidence may be evaluated jointly or in groups, rather than item by item, particularly when pieces of evidence are logically or explanatorily dependent. There is no requirement that each item of evidence be assessed separately.

\item Combine the likelihood ratio with the prior odds to obtain the posterior odds, or relative probability, of one explanation versus the other. In many cases, a prior odds ratio of one may be appropriate, in which case the posterior odds are determined entirely by the evidence. Information often classified as ``prior'', such as base-rate information, may alternatively be treated as part of the evidential body and evaluated through the likelihood, provided it is relevant and admissible.
\end{enumerate}

This procedure does not prescribe how triers of fact actually reason, nor does it require explicit numerical calculation. Rather, it provides a normative algorithmic benchmark for comparative evidential reasoning. When fact-finders aim to determine which explanation better accounts for the evidence, and by how much, posterior odds supply a coherent method for realizing the computational task described by relative plausibility theory.

\section{Algorithmic accommodation of explanatory evidential factors}
\label{sec:evidentialfactors}

A central claim of relative plausibility theory is that it uniquely accommodates explanatory features of evidence evaluation that probabilistic approaches allegedly cannot, including consistency, coherence, fit with background knowledge, simplicity, absence of evidential gaps, and the number or implausibility of assumptions required by an explanation \citep[p.~16]{Allen2019}. Once the distinction between computational and algorithmic levels of description is kept in view, however, this contrast dissolves. Each of these factors can be incorporated within a probabilistic framework as an algorithmic refinement of the comparative task described by RPT. Note that the below algorithmic specifications are merely suggestions for how things could be implemented, no claims are made that these are the only, or even the best, specifications.

\subsection{Background knowledge}

Background knowledge, denoted $K$, has always played a central role in Bayesian inference, even when omitted from notation for simplicity. Likelihood ratios are properly understood as conditional on such background knowledge:
\begin{equation*}
  LR = \dfrac{P(E \mid H_P, K)}{P(E \mid H_D, K)}.
\end{equation*}

Here, $K$ includes general assumptions and beliefs such as ``having a motive increases the likelihood of guilt,'' ``witnesses of dubious character are less credible,'' or ``individuals typically attempt to conceal criminal behavior.'' Background knowledge need not be accurate, uniform across fact-finders, or even well justified; it may be shaped by experience, culture, or popular media. What matters for the present purpose is that such knowledge is explicitly acknowledged as conditioning evidential assessments.

Stipulations of fact are naturally included within $K$. These are propositions accepted as true for the purposes of the litigation and therefore treated as fixed background conditions rather than as items of evidence to be evaluated. Contrary to some claims, this presents no difficulty for probabilistic reasoning \citep{Allen2023}.

\subsection{Simplicity and unlikely assumptions}

Both legal and scientific reasoning exhibit a preference for explanations that are simpler, less \textit{ad hoc}, and require fewer or less implausible assumptions. Within a probabilistic framework, such preferences can be incorporated through what is commonly referred to as an \emph{Occam penalty} -- a factor that down-weights more complex explanations \citep{MacKay2003,Fairfield2022}.

Consider a case in which the prosecution's hypothesis $H_P$ is that the accused alone committed the murder, while the defence's hypothesis $H_D$ posits that six mutually unconnected individuals jointly committed the crime. Suppose further that the available evidence $E$ is equally compatible with both explanations:
\begin{equation*}
  LR = \dfrac{P(E \mid H_P)}{P(E \mid H_D)} = 1.
\end{equation*}

Despite this equality, the defence hypothesis may reasonably be regarded as less plausible due to the number and implausibility of the assumptions it requires. This can be reflected algorithmically by applying a penalty factor $1/F$ to $H_D$:
\begin{eqnarray*}
  LR &=& \dfrac{P(E \mid H_P)}{P(E \mid H_D)} \times \dfrac{1}{1/F} \\
     &=& \dfrac{P(E \mid H_P)}{P(E \mid H_D)} \times \dfrac{F}{1}.
\end{eqnarray*}

After accounting for complexity, the relative support for $H_P$ becomes $F$. The choice of $F$ is an algorithmic judgment, no different in principle from judgments about the strength of motive or credibility of testimony. If one wished to be more explicit, the six individuals posited by $H_D$ generate $\frac{6(6-1)}{2} = 15$ pairwise interactions, which could serve as a lower bound on the penalty.

When explanations are specified using parameterized statistical models, complexity penalties arise naturally through prior distributions on the parameters. Broad priors that permit a wide range of outcomes impose a larger penalty than more constrained priors, reflecting the greater flexibility of the model \citep{MacKay1992,MacKay2003}. Unlikely assumptions can be treated similarly, with lower prior plausibility translating into stronger penalties.

\subsection{Consilience, coherence, and consistency}

Another explanatory consideration emphasized by RPT is how well pieces of evidence fit together under a given hypothesis. Probabilistic frameworks readily accommodate such considerations by allowing evidence to be evaluated jointly rather than in isolation.

Suppose that an eyewitness testimony $E_1$, if true, would exonerate the defendant, but is sufficiently unusual to raise suspicions of fabrication. If a second independent eyewitness provides a similar unusual testimony $E_2$, the combined evidential force may exceed the sum of the individual contributions. In such cases, it is appropriate to evaluate the evidence jointly:

\begin{equation*}
  LR = \dfrac{P(E_1 E_2 \mid H_P)}{P(E_1 E_2 \mid H_D)}.
\end{equation*}

If both witnesses were lying, the probability that they independently fabricate the same unusual story may be very small, leading their testimonies to mutually reinforce one another. Nothing in probability theory requires evidence to be assessed item by item, and joint evaluation can also help to avoid ``double-counting'' when pieces of evidence are logically or explanatorily related \citep{Dahlman2023,Bovens2003,Fairfield2022}.

\subsection{Evidential coverage and completeness}

Finally, critics have argued that probabilistic approaches cannot adequately account for evidential gaps or the incompleteness of the available information \citep{Cohen1977,Tecuci2016}. Consider a case in which the available evidence strongly supports the guilt of the accused -- motive, opportunity, threats, purchase of a weapon, and physical presence -- yet the victim's body has not been found. The absence of this crucial piece of evidence leaves the inference ``out on an inferential limb,'' since the victim may not, in fact, be dead.

Such concerns can be incorporated algorithmically by down-weighting the evidential impact of the available information. One simple method is to raise the likelihood ratio to a power $c \in (0,1]$:

\begin{equation*}
  LR = \left[\dfrac{P(E \mid H_P)}{P(E \mid H_D)}\right]^c.
\end{equation*}

Power likelihoods and related techniques are widely used to discount incomplete, misspecified, or unreliable information, and to increase robustness to outliers or biased judgments \citep{Chen2000,Ibrahim2003,Holmes2017,Aitchison2021,Ghosh2015,Bissiri2016,Lyddon2019,Matsumori2018}. The parameter $c$ may be interpreted as reflecting the degree of evidential coverage or relevance. A value of $c=1$ corresponds to complete evidence, while smaller values indicate increasing incompleteness.

Determining an appropriate value for $c$ may be challenging, but this difficulty concerns algorithmic specification rather than conceptual possibility. The key point is that evidential gaps and incompleteness do not lie beyond the reach of probabilistic reasoning. They can be accommodated within an algorithmic framework that remains fully compatible with the computational task described by relative plausibility theory.

\section{Clarifying alleged objections to probabilistic approaches}
\label{sec:clarifications}

The preceding section showed that the explanatory considerations emphasized by relative plausibility theory can be accommodated within probabilistic reasoning at the algorithmic level. We now return to several objections that continue to motivate resistance to this conclusion. Much of the resistance to probabilistic approaches in the literature stems from a set of recurring objections articulated most prominently by \citet{Allen2019}. When viewed through the lens developed in the preceding sections, these objections do not identify substantive limitations of probabilistic reasoning. Rather, they reflect the difficulties inherent in moving from a high-level computational characterization of juridical proof to a fully specified algorithmic account. Once this distinction is kept in view, the force of the objections is substantially diminished. In addition, nothing in the present argument denies that RPT offers a more natural vocabulary for describing how legal actors talk about evidence.

Allen and Pardo argue that relative plausibility theory avoids four ``problems'' that beset probabilistic approaches \citep[p.~19]{Allen2019}. The first concerns the assignment of numbers. Probabilistic methods, they claim, require priors and likelihood ratios that are ``literally just made up by the decision maker.''~\citep[p.~33]{Allen2019}. By contrast, RPT is presented as an explanatory account for which ``there is no need to quantify the evidence or to attach numbers to the likelihood of legal elements being met.'' \citep[p.~17]{Allen2019}.

This contrast is best understood as a difference in level of description rather than a substantive advantage of RPT. Relative plausibility theory deliberately operates at a computational level, characterizing the inferential task without specifying how it is carried out. As such, it need not confront questions of numerical representation or aggregation. Probabilistic approaches, by contrast, operate at the algorithmic level, where such questions cannot be avoided. The absence of explicit numerical assignments in RPT therefore reflects an incomplete specification of the inferential process, not the resolution of a problem.

Human reasoning routinely exhibits this structure. Individuals can successfully perform complex tasks, such as riding a bicycle, without being able to articulate the underlying physical principles governing their behavior. Describing the task in terms of balance, steering, and pedaling does not render Newtonian mechanics irrelevant; it simply operates at a different level of abstraction. Likewise, the fact that fact-finders may not be explicitly able to assign numerical values does not undermine the relevance of probabilistic accounts as algorithmic descriptions of evidential reasoning.

A closely related concern involves the choice of numerical thresholds corresponding to legal standards of proof, such as the presumption of innocence, proof beyond a reasonable doubt, or the preponderance of evidence. Once again, this concern arises from conflating levels of analysis. A computational-level account need only specify that such thresholds exist and that they play a role in decision-making; it need not fix their numerical values. Determining how such thresholds should be operationalized is an algorithmic and institutional matter, not a conceptual defect of probabilistic reasoning.

The second alleged problem concerns psychological fit. Allen and Pardo argue that fact-finders evaluate evidence holistically rather than item by item, and that probabilistic approaches mischaracterize this process \citep{Allen2019}. This objection again targets particular algorithmic implementations rather than probabilistic reasoning per se. Nothing in probability theory requires evidence to be assessed one item at a time. Evidence may be evaluated jointly, in groups, or as a whole, particularly when pieces of evidence are logically or explanatorily dependent \citep{Fairfield2022}. For example, it may be appropriate to assess an eyewitness's testimony together with considerations of credibility and competence, $P(E_1 E_2 E_3 \mid H)$, rather than treating these factors independently. Probabilistic frameworks are fully compatible with holistic evaluation.

The third alleged problem is the so-called conjunction paradox. In its standard formulation, the problem arises when multiple claims of a civil trial must each be established on the balance of probabilities ($p > 0.5$). If two claims are each supported with probability $0.7$, their joint probability is $ 0.7 \times 0.7 = 0.49$ (assuming independence), which falls below the $0.5$ threshold. However, it is unclear why the joint probability should be computed at all, since the requirement is that each claim individually meets the standard of proof. The choice of legal standards is a policy matter external to both probability theory and RPT. Probability theory can describe the consequences of these standards, but it cannot explain why they take the form they do\footnote{\citet{Kaye1979} made a similar point.}. Much like probability theory can be used to calculate the likelihood of obtaining various hands in Blackjack, it cannot explain why a score of 21 is important.

We need not worry that legal standards are incoherent because the joint probability is less than 0.5. When evidential support is assessed comparatively using relative probabilities, the supposed paradox disappears \citep{Aitken2022,Aitken2024}. Using posterior odds as in Eq.~\eqref{eq:BF}, the combined support becomes $(0.7/0.3) \times (0.7/0.3) = 5.4$, favoring the plaintiff and in line with intuition. The conjunction problem thus reflects a misunderstanding of how probabilistic reasoning is applied in legal contexts, not a flaw in probabilistic inference itself, and RPT provided a clue as to how the problem should be solved.

The final alleged problem concerns the relationship between evidential reasoning and the policy goals underlying standards of proof. Allen and Pardo argue that the explicitly comparative structure of RPT better aligns with these goals \citep{Allen2019}. As shown above, however, probabilistic approaches -- when formulated in terms of posterior odds -- are equally comparative. The difference is not one of structure, but of representation. Relative plausibility theory specifies the comparative task at a computational level, while probabilistic approaches provide one coherent algorithmic realization of that task.

Taken together, these clarifications reinforce the central claim of this paper. The objections commonly raised against probabilistic approaches do not demonstrate their inadequacy as accounts of juridical proof. Rather, they reflect a level-of-analysis error: criticisms appropriate to particular algorithmic implementations do not thereby indirectly support the computational characterization of evidential reasoning as a competing model.

\section{Conclusion}

Debates between relative plausibility theory and probabilistic approaches to juridical proof have often been framed as disputes between competing theories of evidential reasoning. This paper has argued that such disputes rest on a conceptual misunderstanding. By distinguishing between computational and algorithmic levels of description, we have shown that relative plausibility theory and probabilistic reasoning address different explanatory questions and are therefore compatible rather than rivals. Recognizing this distinction clarifies why empirical studies of jury reasoning cannot adjudicate between RPT and probabilistic approaches, and why normative debates about standards of proof should not be framed as disputes over formal representation. Some may not be convinced that RPT and and probabilism are really the same thing at different levels of description. To address this, let us follow David Deutsch's aphorism: ``If you can't program it, you haven't understood it'' \citep[Ch.~7]{Deutsch2012}. Skeptics are invited to write a program that does relative plausible reasoning.

Welch has observed that ``conditions for reconciling these approaches are not yet in place.''~\citep{Welch2020}. We suggest that the reconciliation becomes possible once the relevant levels of analysis are clearly distinguished. When this distinction is kept in view, many familiar objections to probabilistic approaches dissolve, and the relationship between explanation-based and probability-based reasoning becomes intelligible. From this perspective, arguing over whether relative plausibility theory or probabilism is the ``better'' theory of juridical proof is misguided. These approaches do not compete to explain the same phenomenon at the same level of description.

We therefore concur with \citet{Schauer2019} that relative plausibility theory does not constitute a paradigm shift in the theory of evidence and proof. Its principal contribution lies in clarifying the comparative and explanatory nature of juridical proof at a high level of abstraction. Probabilistic approaches do not undermine this insight; rather, they complement it by specifying how such comparative judgments can be represented and combined in a coherent manner. Allan and Pardo deserve credit for clearly articulating the computational level description of juridical proof, for many research and legal questions are probably better understood at the computational level. And an argument can be made that (too) much legal reasoning was happening at the algorithmic level when the computational level is more appropriate. In addition, they rightly noted that evaluating evidence is a comparative process.

Progress in evidence scholarship depends on resisting the temptation to frame computational and algorithmic accounts as competitors. Once the relevant levels of analysis are properly distinguished, relative plausibility theory and probabilistic reasoning can be seen as parts of a unified research program aimed at understanding how evidence should be assessed, compared, and acted upon in legal contexts.

\section{Declaration of conflicting interests}

The author(s) declared no potential conflicts of interest with respect to the research, authorship, and/or publication of this article.

\section{Funding}
The author(s) received no financial support for the research, authorship, and/or publication of this article.

\end{document}